\documentclass[prb,twocolumn,showpacs,superscriptaddress,footinbib]{revtex4}


\newcommand{\beq}{\begin{equation}}
\newcommand{\eeq}{\end{equation}}
\newcommand{\beqn}{\begin{eqnarray}}
\newcommand{\eeqn}{\end{eqnarray}}

\newcommand{\slp}{\raise.15ex\hbox{$/$}\kern-.57em\hbox{$ \partial $}}
\newcommand{\lnA}{\raise.15ex\hbox{$/$}\kern-.57em\hbox{$A$}}

\usepackage{graphicx}
\usepackage{epsfig}
\usepackage{bm}

\begin{document}
\title{Backscattering off a
dynamical impurity in 1D Fermi systems: A perturbative
computation}

\author{Carlos M.\ Na\'on}
\author{Mariano J.\ Salvay}
\affiliation{Departamento de F\'\i sica, Facultad de Ciencias
Exactas, Universidad Nacional de La Plata, CC 67, 1900 La Plata,
Argentina} \affiliation{Instituto de F\'\i sica La Plata, Consejo
Nacional de Investigaciones Cient\'\i ficas y T\'ecnicas,
Argentina}

\date{October 9, 2006}

\begin{abstract}
We investigate the problem of backscattering off a time-dependent
and spatially extended barrier in a one-dimensional electron gas.
By performing a perturbative expansion in the backscattering
amplitude, we compute the total energy density of the system. We
show how the free fermion spectrum and the conductance of the
system are affected by the interplay between dynamical and
geometrical properties of the impurity.
\end{abstract}

\maketitle

The physics of tunneling through time-dependent barriers is a
topic of great importance in the subject of correlated quasi-one
dimensional electron transport. A considerable effort has been
made over the last few years, focused on the understanding of
quantum transport, both experimentally \cite{experiments} and
theoretically \cite{theories}. Despite the possible technological
advances that the control of charge and spin currents could bring
up, we are faced with the fundamental question of what we could
learn about electron correlations using time-dependent potentials
as out-of-equilibrium probes. To answer this question, at least
partially, more theoretical work is needed to understand the
detailed dynamics induced by this type of perturbation.

An interesting observable that characterizes a tunneling process
is the energy resolved current $j(\omega)$. In Ref.
\onlinecite{Gogolin}, the relevance of this quantity in the upper
region of the spectrum (i.e., above the Fermi energy, $\omega >
E_F$) was emphasized, in connection to correlations in the leads.
To leading order in the tunneling amplitude, the energy resolved
current $j(\omega)$ is related to a simpler observable, namely,
the electron energy distribution function $n(\omega)$, also
referred to as the total energy density (TED) in the literature
\cite{Palmer} (for a precise mathematical definition see Eq.
(\ref{d}) below). This quantity describes the perturbation of the
ground state in the leads due to tunneling processes. For a
non-correlated material, for instance, $n(\omega)$ should vanish
above the Fermi surface. Any population of the spectrum above the
Fermi energy is originated from a combined effect of correlations
and multi-particle tunneling, due, on its turn, to out of
equilibrium processes \cite{Gogolin}. In Ref. \onlinecite{Komnik},
$n(\omega)$ was evaluated in a model of correlated one-dimensional
fermions with a time-dependent impurity coupled to the electron
density through a forward-scattering coupling. In Ref.
\onlinecite{Nos} this model was analyzed by means of functional
bosonization \cite{Nos1}, focusing, in particular, on the
transients produced by turning on the oscillatory impurity
strength.

It is important to notice, however, that backscattering effects
are expected to be relevant as a rule, in all but rather
exceptional experimental settings \cite{Poncharal}. Possible
experimental realizations where backscattering will play a central
role are 1D wires in the presence of a time-dependent gate voltage
and a Hall bar with a constriction \cite{Milliken}. The problem of
backscattering by dynamical impurities is usually a very difficult
one. Some models, like the spinless Luttinger model with a
delta-like impurity, can be solved exactly for the specific value
of the Luttinger parameter $K=\frac{1}{2}$ \cite{Chamon3}.
However, for general strengths of the electron-electron
interaction and finite-ranged impurities, there are no available
closed analytical solutions (even in the free case). Therefore, it
is of fundamental importance to develop different strategies in
such a context. Recently \cite{Gefen}, effects of backscattering
in a Luttinger liquid due to a time-dependent ultralocalized
impurity were studied perturbatively, finding a striking
enhancement of the total current for special values of the
Luttinger parameters. An alternative, non-perturbative point of
view was adopted in Ref. \onlinecite{Barci}, where an adiabatic
approximation was invoked in order to get the distortion of the
non correlated TED due to the backscattering amplitude and the
geometry of the impurity, i.e. for free fermions in the presence
of an extended barrier. Since the main results of Ref.
\onlinecite{Barci} (a peak structure of the TED) were obtained in
a strong coupling and low frequency regime ($\mid \omega \mid \gg
\Omega $), it is certainly desirable to have a quantitative
knowledge of the TED for the same problem (free fermions with a
time-dependent barrier) but in the weak coupling regime and for
all external frequencies. This is the main motivation for the
present work. We study the effects that the backscattering off an
extended dynamical impurity of width $a$ and amplitude $g_b$,
oscillating with frequency $\Omega$, will have on the spectrum of
a one-dimensional fermion gas. Our results, though obviously valid
only for $\frac{g_b a}{\hbar v_F}$ sufficiently small, are not
restricted to small values of the external frequency. We then
expect to capture the main features related to the time-dependent
nature of the perturbation. We compute the TED up to second order
in the backscattering parameter $\frac{g_b a}{\hbar v_F}$. We also
evaluate the change in the conductance $\Delta G$ produced by the
time-dependent barrier. We show that, in contrast to the result
obtained in Ref. \onlinecite{Gefen} for a Luttinger system with an
ultralocalized impurity (when specializing the result to the case
of noninteracting electrons), in a system of non correlated
electrons an extended geometry gives rise to a non trivial
dependence of $\Delta G$ on the frequency of the perturbation. In
particular we find that the conductance of the system remains
unchanged for high frequencies ($\frac{\Omega a}{ v_F}\gg 1$).

As the computational starting point, let us consider the following Hamiltonian, which
describes the interaction of spinless fermions with an external effective time-dependent
potential $V(x,t)$, responsible for backscattering transitions between right and left movers:

\beq H=H_0+H_{\rm imp} \ , \ \label{H} \eeq where \beq H_0=i\;
\hbar \; v_F  \int dx\;\; \left(\psi_R^{\dagger}
\partial_x\psi_R-\psi_L^{\dagger}\partial_x\psi_L\right) \ , \
\label{H_0} \eeq and \beq H_{\rm imp}=g_{b}\int
dx\;\;\left(\psi^{\dagger}_R\psi_L+\psi^{\dagger}_L\psi_R\right)\;V(x,t)
\ . \ \label{H_b} \eeq Above, $g_{b}$ is the coupling constant
associated to the backward scattering of electrons caused by the
presence of a time-dependent harmonic barrier. Let us mention here
that the simultaneous presence of both forward and backward
scattering by the impurity does not bring about any new effect, at
least up to second order in the couplings. In fact, in the absence
of impurity-backscattering the TED has been computed exactly in
references \onlinecite{Komnik} and \onlinecite{Nos}, showing a
sideband structure that reflects the inelastic nature of
time-dependent scattering. Since up to second order in the
couplings the crossed term that would relate forward and backward
contributions vanish, the combined effect will be a direct
superposition of the above mentioned results and the ones we
present here.

Although we have verified that our method works independently of
the explicit details of $V(x,t)$, in order to explore the effect
of finite range barriers, we consider a square potential profile,

\beq
V(x,t)=(\Theta (x + a/2) - \Theta(x - a/2))\cos(\Omega t) \ , \
\label{V}
\eeq
where $a$ is the width of the square potential and $\Omega$ the oscillation frequency.

We are particularly interested in obtaining the TED for the above model. We recall that in
the Wigner representation the TED can be written in terms of the fermion correlation function
as
\begin{equation}
n(\omega,X,T) = -i \int^{\infty}_{-\infty} d\tau\; e^{i \omega
\tau}G_{+ -}(r = 0, X, \tau, T)  \ , \ \label{d}
\end{equation}
where we have introduced the closed time path formalism \cite{CTP}
in which fermion propagators are time-ordered along the usual
Schwinger-Keldysh time contour:\begin{eqnarray} G_{+
-}(\textbf{x}, \textbf{x}') & = & ~~i \langle
\Psi^{\dag}(\textbf{x}')\Psi(
\textbf{x})\rangle , \nonumber \\
G_{- +}(\textbf{x}, \textbf{x}') & = & -i \langle
\Psi(\textbf{x})\Psi^{\dag}(\textbf{x}')\rangle , \nonumber
\\ G_{+ +}(\textbf{x}, \textbf{x}') & = & -i \langle T \Psi(\textbf{x})\Psi^{\dag}( \textbf{x}')\rangle ,
\nonumber \\ G_{- -}(\textbf{x}, \textbf{x}') & = & -i \langle
\widetilde{T}\Psi(\textbf{x})\Psi^{\dag}( \textbf{x}')\rangle ,
\label{masmenos}\end{eqnarray} where $T$ and $\widetilde{T}$
denote the time and anti-time ordering operations respectively.
Above, $r, \tau$ and $X, T$ are the spatial and temporal relative
and center of mass coordinates, respectively.

In realistic systems the frequency $\Omega$ is expected to be
quite high so that it is unlikely that the explicit time
resolution of the TED would be experimentally accessible. Then, it
is natural to consider the time average, over the period of the
perturbation:
\begin{equation}\overline{n}(\omega,X) =
\frac{\Omega}{2 \pi} \int _{0}^{2 \pi / \Omega}  d T \, n(\omega,
X, T).\label{h}\end{equation} Let us stress that this averaged TED
is a purely dynamical quantity, i.e. in general it is not
connected to the static case. In particular, as explained in
Ref.\onlinecite{Barci} the static limit $\Omega \rightarrow 0$
cannot be reproduced from this expression (see below).

At this point we calculate the average TED as an expansion up to
the first non trivial order in the dimensionless parameter $g_{b}
a/\hbar v_F$. As it it is usual in the context of 1D fermionic
models, we work in the chiral representation, introducing a spinor
$\Psi$ with components $\psi_L$ and $\psi_R$. Computing the
corrections to the Green functions $G^{R}_{+-}$ and $G^{L}_{+-}$
due to the time-dependent barrier, one readily verifies that the
first order contributions vanish. Considering then the second
order corrections, inserting the results in (\ref{d}) and finally
using the definition given in (\ref{h}), we obtain the following
expressions for the average TED's for both right and left
components ($\overline{n_{R}}(\omega,X)=
\overline{n_{L}}(\omega,-X)=N(\omega,X)$):

\begin{widetext}\begin{eqnarray} N(\omega,X) = \frac{1}{v_F}(\Theta(-\omega) - [ \Theta(X + a/2) - \Theta(X - a/2) ]\,
\{\Theta(-\omega) \, \frac{g_{b}^{2}a^{2}}{2\hbar^2v_F^2} \,
(F((2 \omega + \Omega)a/v_F) + F((2 \omega - \Omega)a/v_F))  \nonumber\\
-\Theta(-\omega) \, \frac{g_{b}^{2}( a/2 - X)^{2}}{2\hbar^2v_F^2}
\, (F((2 \omega + \Omega)(a/2 - X)/v_F) + F((2 \omega -
\Omega)(a/2 - X)/v_F)) \nonumber\\- \Theta(-\omega - \Omega) \,
\frac{g_{b}^{2}(a/2 + X)^{2}}{2\hbar^2v_F^2} \, F((2 \omega +
\Omega)(a/2 + X)/v_F) - \Theta(-\omega + \Omega) \,
\frac{g_{b}^{2}(a/2 + X)^{2}}{2\hbar^2v_F^2} \, F((2 \omega - \Omega)(a/2 + X)/v_F) \} \nonumber\\
- \Theta( X  - a/2 )\, \frac{g_{b}^{2}a^{2}}{2\hbar^2v_F^2} \, \{
[ \Theta(-\omega) - \Theta(-\omega - \Omega) ]F((2 \omega +
\Omega)a/v_F) + [ \Theta(-\omega) - \Theta(-\omega + \Omega) ]F((2
\omega - \Omega)a/v_F)\}),\label{ine}
\end{eqnarray}\end{widetext}
where $F(z) = \frac{ 1 - \cos{z} }{z^2}$. Let us recall that this
result is valid for $| \frac{g_{b} a}{\hbar v_F} | \ll 1$.
Analyzing the above expression one sees that the average TED is a
superposition of free TED's centered at the origin of energies and
at $\pm \Omega$. This feature is a direct consequence of the
perturbative order we are working at. Indeed, we have verified
that terms corresponding to larger shifts in $\omega$ will also
contribute to $N(\omega)$ when next order corrections are taken
into account. In analogy to the behavior of the TED when only
forward scattering barriers are present \cite{Nos1}, the
coefficients that weight the contributions of the various free
TED's depend on $\Omega$ and the impurity geometry through the
width $a$. However, in contrast to that case, in which the average
TED coincides with the static case (which is in turn equal to the
free value: $N(\omega) = \Theta (- \omega)$) {\em at every spatial
point} when $\Omega \rightarrow 0$, in the present case this limit
cannot be obtained {\em inside the barrier}. In fact, one gets one
half of the right answer for the static limit in this perturbative
order. On the other hand, outside the barrier there is no problem
and both results for $N(\omega)$ coincide for $\Omega \rightarrow
0$ (these results are also equal to the free case, see
(\ref{ine})). This disagreement found for TED's evaluated inside
the barrier is not surprising since the limit $\Omega \rightarrow
0$ in Eq. (\ref{h}) is not well defined.
\begin{figure}\begin{center}
\hspace{-1.0cm} ~~~~~\includegraphics[width=5.4cm,
height=4.3cm]{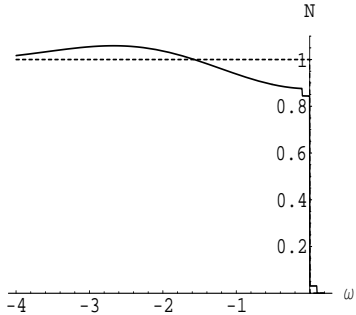} \caption{Averaged TED for $g_{b} =
1/\sqrt{2}$, $a = 1$ and $\Omega = 0.1$ ($\hbar=v_F=1$). The
dashed line corresponds to the free case.} \label{fig1}
\end{center}\end{figure}
\begin{figure}\begin{center}
\hspace{-1.0cm} ~~~~~\includegraphics[width=5.4cm,
height=4.3cm]{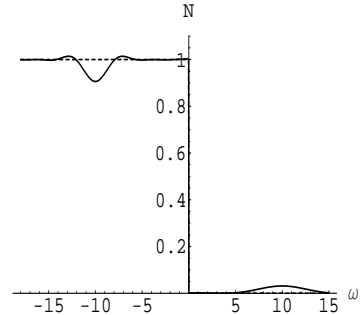} \caption{Averaged TED for $g_{b} =
1/\sqrt{2}$, $a = 1$ and $\Omega = 20$ ($\hbar=v_F=1$).}
\label{fig2}
\end{center}\end{figure}

In figures 1 and 2 we display the behavior of
$\overline{n}(\omega)=N(\omega)$ at the center of the barrier and
for fixed values of $g_{b}$ and $a$. We can identify two different
regimes in the behavior of $N(\omega)$, according to the values of
$\frac{\Omega a}{v_F}$. For $\frac{\Omega a}{v_F} \ll 2.67 $
(please see figure $1$), $N(\omega)$ has a maximum in $-
\frac{\omega a}{v_F} \approx 2.67$; this corresponds to the {\em
quasi-static} region. Figure $2$ shows an example of the other
regime, the high-frequency region, where $\frac{\Omega a}{v_F} \gg
2.67 $. Here we note the appearance of a depression in the
spectrum centered at $\omega = - \Omega/2$, and a peak at $\omega
= \Omega/2$. The peculiar value $\frac{\Omega a}{v_F} \approx 2.67
$ has been determined numerically, searching for the region of
maximum superposition of both previous effects. It is interesting
to note that one peak centered around this value also appears in
the context of the adiabatic approximation, when considering the
case $g_b \,a =1$ (figures 1 and 2 of Ref.\onlinecite{Barci}).
However we do not take the comparison further, since the adiabatic
approach of Ref. \onlinecite{Barci} focused on the strong coupling
region, whereas here we deal with the opposite regime.

In contradistinction to the case of forward scattering barriers,
impurities of the backscattering type affect the transport
properties of the system. In order to obtain the conductance $G$
for this system, we must analyze the linear response of the
current under the influence of an external bias $V$. The effect of
this voltage can be introduced, as usual, by modifying the
Hamiltonian density ${\cal H}$ given in (\ref{H}) as ${\cal H}
\rightarrow {\cal H} + \mu_{L} \psi_L^{\dagger}\psi_L $, where
$\mu{_L} = -e V$ is the chemical potential coupled to fermions of
left chirality. The conductance is then given by
\begin{equation} G = lim  _{V
\rightarrow 0} \frac{J}{V},\end{equation}with
\begin{equation}J= -i e v_F\frac{\Omega}{2 \pi} \int ^{\frac{2 \pi}{
\Omega}}_{0}[G_{+-}^{R}(x , t, x, t) - G_{+-}^{L}(x , t, x, t)] dt
,\end{equation} where $J$ is the dc component of the current. As
is well known, in the absence of impurities, the conductance of a
$1 D$ non correlated Fermi system is $ G = e^{2}/h $ , where h is
Planck's constant. We have computed the second order correction to
$G$ due to the presence of the time-dependent backscattering
barrier considered in this work. The result is
\begin{equation}\label{conductance}\Delta G = - \frac{e^{2}}{h} \, 2 \, \frac{g_{b}^{2}}{\hbar^{2}}\,\frac{\sin ^{2}[\Omega a
/2 v_{F}]}{\Omega^{2}}.\end{equation}

We then found that the conductance decreases for $\Omega a /2
v_{F}\neq n \pi$. The magnitude of the effect depends on the
amplitude of the barrier, its width and the frequency of the
oscillation. We see that in contrast to the result obtained for a
point impurity \cite{Gefen}, where $\Delta G$ is independent of
frequency, we get a non trivial oscillatory behavior of $\Delta G$
as function of the external frequency $\Omega$. In particular, in
the low frequency regime we predict a maximum decrease of $G$,
similar to the behavior corresponding to static barriers, as
expected. On the other hand, for $\Omega a /2 v_{F}= n \pi$ (with
$n$ an integer greater than 1) $G$ is not affected by the barrier.
The same phenomenon takes place in the high frequency region.

\vspace{1cm}

To summarize, we have studied the effect of a backscattering
time-dependent barrier on the spectrum and transport properties of
a non correlated 1D electronic system. We focused our attention
not only on the out of equilibrium physics caused by a dynamical
impurity but also on the role played by its extended geometry. We
performed a perturbative computation of the total energy density
of the system $N(\omega)$. In the low frequency regime we found a
maximum of $N(\omega)$ in $- \frac{\omega a}{v_F} \approx 2.67$;
this corresponds to the {\em quasi-static} region. In the high
frequency regime $N(\omega)$ displays a depression in the spectrum
centered at $\omega = - \Omega/2$, and a peak at $\omega =
\Omega/2$. Concerning the conductance of the system we showed
that, in contrast to the behavior predicted for an ultralocalized
barrier, for an extended impurity it changes as an oscillatory
function of $\Omega a /2 v_{F}$ (see equation
(\ref{conductance})). This result, together with the one obtained
in reference \onlinecite{Gefen}, could be used to experimentally
characterize the spatial structure of constrictions through
conductance measurements.

\acknowledgments

This work was supported by Universidad Nacional de La Plata and
Consejo Nacional de Investigaciones Cient\' ificas y T\'ecnicas,
CONICET (Argentina).

\end{document}